\documentclass[12pt]{article}
\usepackage{epsfig}
\usepackage{graphicx}
\newcommand{\stau}{$\tilde \tau~$}

\newcommand{\tanb}{$tan \beta~$}
\newcommand{\tanbz}{$tan \beta$}
\newcommand{\kapa}{$\kappa$}
\newcommand{\none}{$\chi_1^0~$}
\newcommand{\nonez}{$\chi_1^0$}

\begin{document}
\setlength{\baselineskip}{.6cm} \setlength{\parskip}{1mm}

\begin{titlepage}
\bigskip
\begin{flushright}
   {2.05.2004\\
      WIC/12/04-MAY-DPP}
\end{flushright}
\bigskip \bigskip
\begin{center}{\LARGE\bf A New Technique for Finding the Underlying Model Parameters in GMSB.}
\end{center}
\bigskip
\begin{center}{Ehud Duchovni \footnote{ehud.duchovni@weizmann.ac.il} and Peter
Renkel \footnote{Renkel@wisemail.weizmann.ac.il} }\end{center}

\begin{center}{\it Particle Physics Department,\\
Weizmann Institute of Science, Rehovot 76100, Israel}
\end{center}

\bigskip\bigskip\bigskip

\begin{abstract}
A new, model dependent way of uncovering some of the values of the
underlying GMSB parameters is proposed.  This method is faster
than the existing procedures and once enough luminosity is
available, can be used to check the consistency of the model.
\end{abstract}
\end{titlepage}
%
%
\newcommand{\qq}         {\mbox{$Q^2$}}

\section{Introduction}
This note assumes that after a short run at LHC one will find out
that nature is supersymmetric and follows the Gauge-Mediated
SuperSymmetry Breaking (GMSB) scheme. At that point one will face
the task of uncovering the value of the underlying model
parameters. In the simplified version of such a model 6 such
parameters exist \cite{Wells}, namely:
$$\Lambda,~M,~Tan \beta,~ N,~ sign (\mu),~ \kappa$$
For simplicity sake it is assumed here that \emph{N=1}. This
assumption gives rise to the lightest possible mass spectrum of
SUSY particles for a given M and $\Lambda$. The case of $N>1$
will be discussed in a future note.\\

\begin{figure}
  \includegraphics[width=12cm]{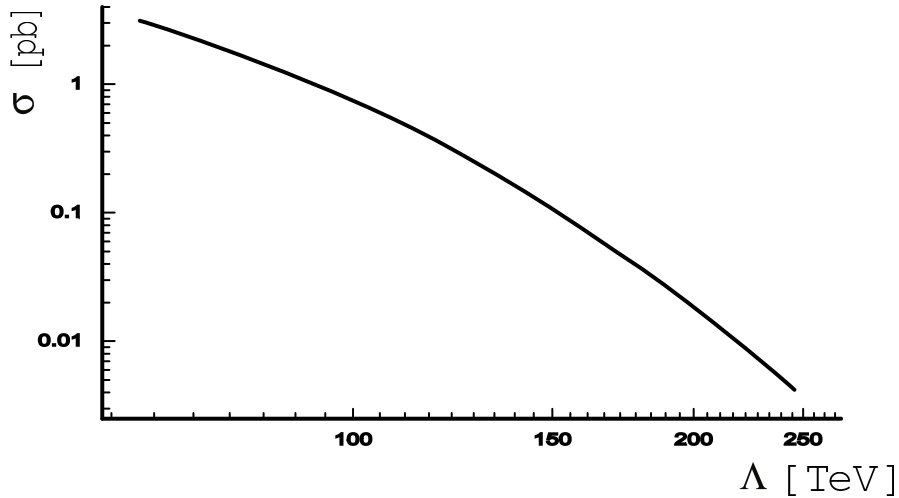}\\
  \caption{Total SUSY cross-section as a function of $\Lambda$
  value.}
  \label{lambda}
\end{figure}

The value of \emph{M} has marginal effect on the phenomenology
which will be discussed below because it enters only through the
logarithmic running of the masses in the renormalization group
equations, and will therefore, be ignored hereafter. The sign of
$\mu$ has also a small effect and will be only briefly treated.
\kapa~ controls the way in which the LSP, namely the Gravitino
couples to the other particles, in particular to the NLSP. It is
assumed here to be small. In case it is large it will be
determined by measuring the lifetime of the NLSP. The value of
$\Lambda$ will be determined directly from the production cross
section of SUSY events (Figure \ref{lambda}) and due to its
triviality will not be further discussed here.\\

The parameter which is at the focus of the present study is \tanb.
\tanb is traditionally \cite{TDR} determined using the mass of the
Higgs boson and to a lesser extent by the mass of the 3rd
generation sparticles. However, if nature is indeed
supersymmetric, the Higgs boson mass is expected to be at most 130
GeV. In this mass range the discovery and the mass determination
of the Higgs boson require relatively high luminosity (the
cross-section times branching ratio of a 130 GeV Higgs boson to a
$ZZ^*$ and later to a 4-lepton final state is only $\approx 3
fb)$. Mass reconstruction is also a non-trivial task. Only a small
fraction of the produced events are suitable for mass
reconstruction. Consequently one will be able to determine \tanb
with a reasonable accuracy only after collecting a significant
amount of data.\\

\begin{figure}
  \includegraphics[width=14cm]{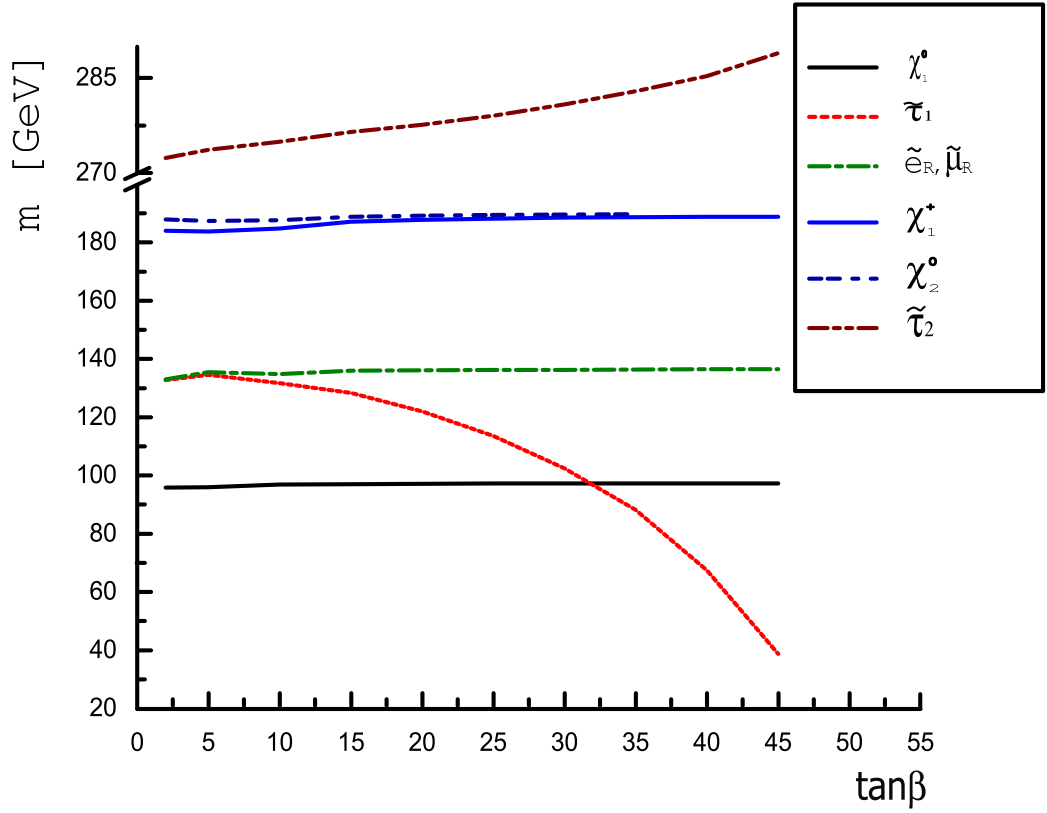}\\
  \caption{Mass dependence on \tanbz. The solid black line represents
  the mass of the \none, the dotted red line that of the \stau lepton,
  the dashed-dotted green line represents the mass of the
  $\tilde e_R$ and $\tilde \mu_R$ which are degenerated, the
  dashed blue line that of the $\chi^{\pm}_1$,
  the upper solid light blue line is for the $\chi^0_2$ and the brown
  dashed-dotted line represents the mass of the $\tilde \tau_2$.
  The masses here are plotted for the case of $\Lambda=75~TeV$,
  M=250 TeV and positive $\mu$.}
  \label{mass}
\end{figure}

The method which is proposed in the present study is based on the
fact that the mass of the \stau lepton depends on the value of
\tanbz. For low \tanb the \stau lepton is relatively heavy
\footnote{since only the $\tilde \tau_1$ lepton is relevant in the
present study the subscript '1' was omitted.} and it decays into
the NLSP, namely to the \nonez. For high values of \tanb the \stau
lepton becomes lighter and eventually it becomes the NLSP (Figure
\ref{mass}). This change in the \stau lepton mass leads to a
change in the average number of $\tilde \tau$ leptons in SUSY
events. The \stau lepton decay gives rise to a final state that
contains a $\tau$ lepton. Hence, it is argued here that one does
not need to reconstruct and measure the mass of the $\tilde \tau$
lepton, rather, \emph{one can determine \tanb by simply counting
the number of $\tau$ leptons as well as electrons and muons in
SUSY events}.

\section{Description of the technique}
One may divide the relevant \tanb interval into 3 distinctive
regions. The first one, the \emph{low} \tanb region, is
characterized by a light \none (whose mass is independent of
\tanbz) which is the NLSP and a relatively heavy \stau lepton
(whose mass depends on \tanbz). In this region one expects to see
a large number of isolated energetic photons which are the result
of the only possible \none decay mode, namely, $\chi_1^0
\rightarrow \gamma \tilde G$. One would also expect to see
relatively sizable number of isolated energetic $\tau$ leptons.
These $\tau$ leptons can originate from two main sources: the
decay $\tilde \tau \rightarrow \chi^0_1 \tau$ which gives rise to
relatively soft $\tau$ leptons due to the small $\tilde \tau -
\chi_1^0$ mass difference, and the decay $\chi^0_2 \rightarrow
\tilde \tau \tau$ which gives rise to a harder spectrum of $\tau$
leptons. The \emph{low} \tanb region extends over a very large
region of \tanb and the determination of the value of \tanb inside
this region will be the main issue of this study.

The other extreme case: the \emph{high} \tanb region, is
characterized by a light \stau lepton which is the NLSP and a
somewhat heavier \nonez. In this region one expects to see a large
number of isolated energetic $\tau$ leptons and no isolated
energetic photons. Here one would have an additional source for
hard $\tau$ leptons, namely, the $\tilde \tau \rightarrow \tau
\tilde G$ decay. This region, in principle, can extend up to \tanb
of about 50. However, for moderate values of $\Lambda$ the mass of
the $\tilde \tau$ lepton drops, for high \tanbz, below the LEP
limit \cite{lep_tau} of 86.9 GeV. Consequently, the range of \tanb
covered by this region is quite limited.

The \emph{intermediate} region is the region in which the mass
difference between the \stau lepton and the \none is smaller than
the mass of the $\tau$ lepton and one cannot decay to the other.
In this region both the number of isolated photons and $\tau$
leptons are 'frozen'. For the case in hand of $\Lambda=75$ TeV
this happens for \tanb between 32.8 and 34.6.

\begin{figure}
  \includegraphics[width=14cm]{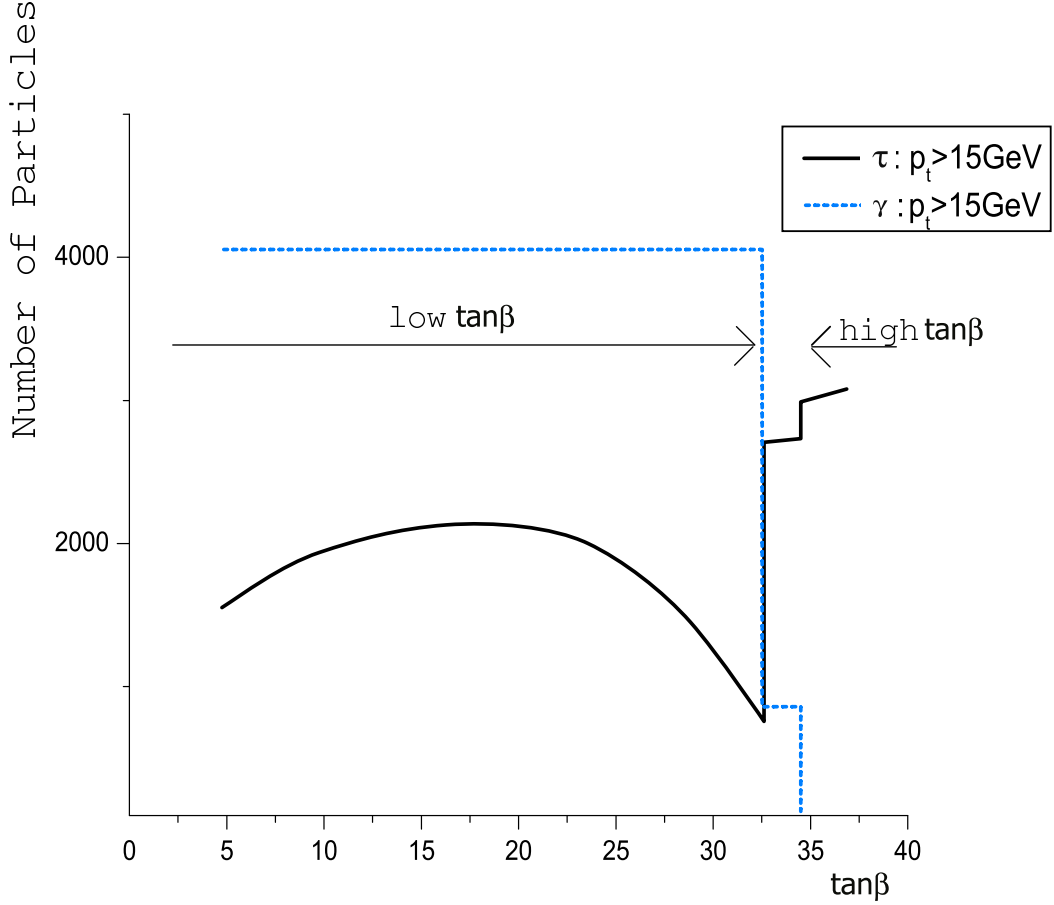}\\
  \caption{The expected average number of
isolated energetic photons and $\tau$ leptons as a function of
\tanb. The number of $\tau$ leptons is shown after applying a
$p_t$ cut of 15 GeV (full black curve) on its visible decay
products (no detector simulation). The number of photons (blue
dotted) was computed with the same $p_t$ cut. The numbers were
computed assuming luminosity of 1 $fb^{-1}$, $\Lambda=75~TeV$,
$M=250~TeV$ and positive $\mu$.}
  \label{average}
\end{figure}

The main features of the events, namely the average number of
isolated energetic photons and $\tau$ leptons as a function of
\tanb is shown in Figure \ref{average} for a machine luminosity of
1 $fb^{-1}$, $\Lambda=75~TeV$ and $M=250~TeV$. The number of
$\tau$ leptons in the figure is computed after applying a $p_t$
cut of 15 GeV on visible $\tau$ lepton decay products and without
any detector simulation. These results show that by a simple
photon and $\tau$ leptons counting experiment one will, in
principle, be able to determine which is the relevant region of
\tanbz: In the low \tanb one would see a large number of photons
and a sizable amount of $\tau$ leptons. In the intermediate region
the number of photons will drop significantly but one will still
be able to see a few hundred photons per $fb^{-1}$. One will see,
in addition, a large number of $\tau$ leptons. In the high \tanb
region one will observe a very large number of $\tau$ leptons and
no isolated photons coming from \none.

\begin{figure}
  \includegraphics[width=14cm]{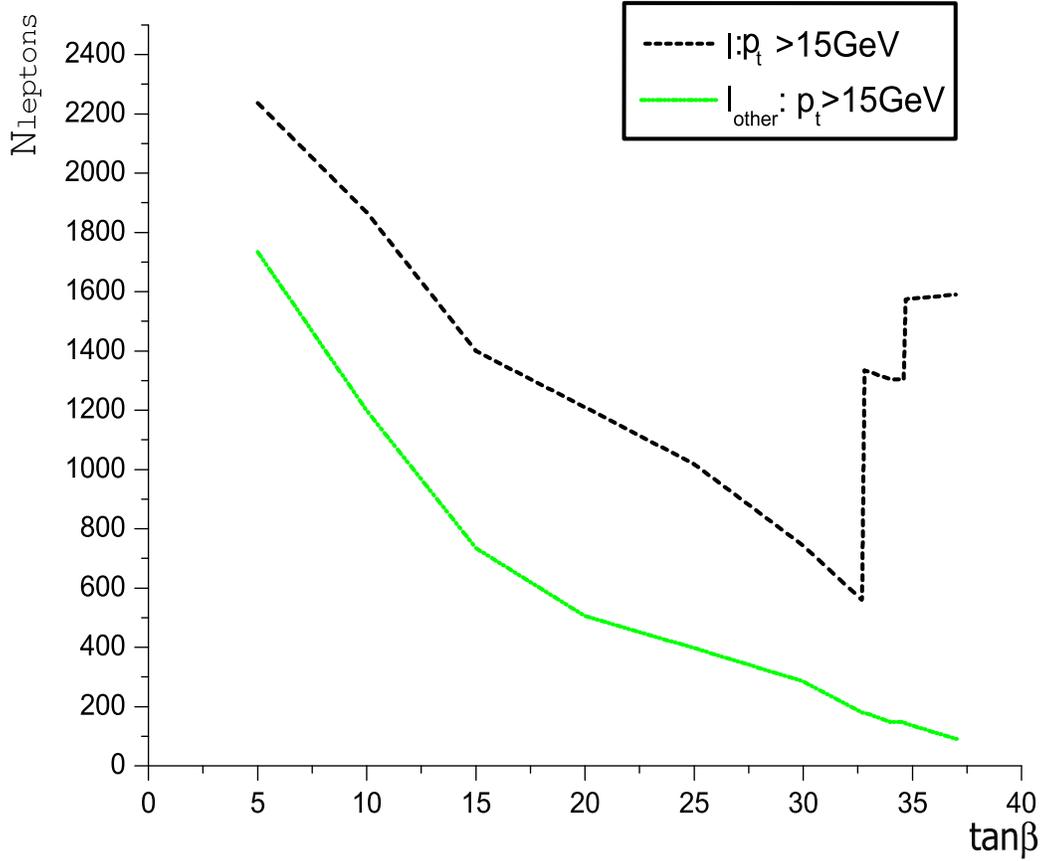}\\
  \caption{The dependence of the number of light leptons
  (electrons and muons) on the value of \tanbz. The black (dashed)
  curve is for the total number of lepton and at high values of
  \tanb is dominated by leptons coming from $\tau$ lepton decays.
   The blue (solid) curve represents the leptons that do not come
   from $\tau$ lepton decays (mostly from slepton and gaugino
   decays). The plot is done with equivalent luminosity of
   1 $fb^{-1}$.}
  \label{nlep}
\end{figure}

As mentioned above, the low \tanb region extends for $\Lambda$=75
TeV and positive $\mu$ from $tan \beta \approx 2$ till $tan \beta
\approx 32.7$. This is a very large region and one would like to
be able to determine \tanb inside this region. The number of
photons is absolutely flat and offers no clue for \tanb
determination, but the number of $\tau$ leptons shows some
dependence on \tanb and a measurement of $N_{\tau}$ will allow one
to determine \tanb value inside this region. The number of $\tau$
leptons increases initially as \tanb increases since heavy
gauginos prefer decaying into $\tilde \tau$ due to the drop in its
mass compared to the mass of the other slepton. This effect is
then balanced and overcome by the effect of the $p_t$ cut of the
$\tau$ lepton, as the resulting $\tau$ leptons at high \tanb
values are softer due to the smaller mass difference between the
$\tilde \tau$ and the $\chi^0_1$. The resulting shape of the
$N_{\tau}$ curve shows that the determination of \tanb in this way
will suffer from a low-high ambiguity. One may exploit the $\tau$
lepton momentum distribution to resolve this ambiguity but a
higher sensitivity to \tanb can be achieved by counting the number
of light leptons, namely, electrons and muons which is shown in
Figure \ref{nlep}. At low values of \tanb the $\tilde \tau$ lepton
is relatively heavy and consequently its production in gaugino
decays is as likely as the production of light leptons. At higher
values of \tanb the $\tilde \tau$ lepton becomes lighter and its
production in those decay is preferred. As a result the number of
prompt light leptons (not from $\tau$ lepton decay) is dropping.
This is clearly shown in Figure \ref{nlep}.

\section{Results Using Fast Simulation}
The fast Atlas detector simulation program was used in order to
estimate the sensitivity of the procedure which was outlined above
under semi-realistic conditions. In addition to the expected
degradation of the sensitivity due to detector simulation the
effects of possible background sources were also taken into
account. Signal events were simulated using the Isajet
\cite{Isajet} MC program. QCD, $W^{\pm}$ and $W^{\pm}$+jets, $Z^0$
and $Z^0$+jets as well as $t \bar t$ events were also simulated
using Isajet and the same detector simulation program. The
isolation of SUSY events was based on the presence of isolated
photons in the final state, namely on two simple requirements:

\begin{enumerate}
\item at least one isolated photon \footnote{an isolated photon is
defined as an electromagnetic cluster with $E_t>15 GeV$ in
$|\eta|<2.5$ range, separated from the closest jet by $\Delta
R>0.4$ and from the closest lepton by $\Delta R>0.15$ and
surrounded by a $\Delta R>0.2$ cone in which less than 10 GeV were
recorded in the calorimeter.} with $p_t>15~GeV$;
\item$E_T^{miss}>100~GeV$.
\end{enumerate}

these simple cuts reduced the SM background to a negligible level.
The hadronic $\tau$ leptons are identified using the Weizmann
algorithm \cite{tauid}. In this algorithm one starts with jets. A
jet is accepted as a $\tau$ if it fulfills the following
requirements:

\begin{enumerate}

\item

there is 1 charged track with $p_{Ttrack} > 5~GeV$ in a cone
$\Delta R < 0.07$, $\Delta R = \sqrt{\Delta \phi^2 + \Delta
\eta^2}$, around the jet axis;

\item

an isolation fraction, i.e. fraction of transverse energy in a
ring $0.2 < \Delta R < 0.4$ around the jet axis, is smaller than
0.1;

\item

$p_T$ of the jet (non-calibrated) is greater then 15 GeV;

\item

number of the hit cells within a jet cone of $\Delta R < 0.4$ is
smaller than six.

\end{enumerate}

\begin{figure}
  \includegraphics[width=15cm]{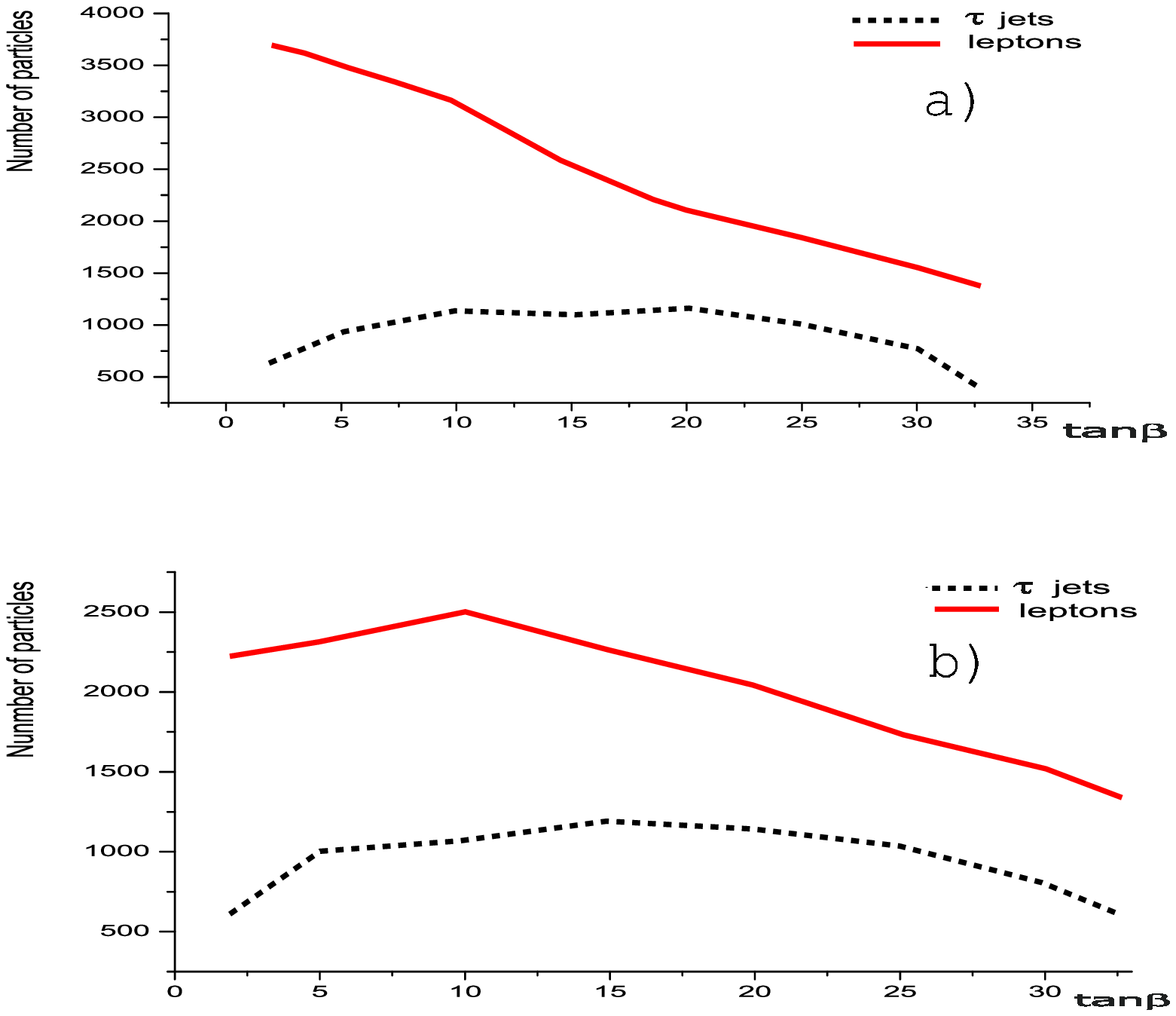}\\
  \caption{The number of expected $\tau$ leptons (dotted black curve)
  and light leptons (red solid curve) for different values
  of \tanb as predicted using fast simulation for a
  a luminosity of 4 $fb^{-1}$. A cut of
  $p_t>15~GeV$ visible transverse momentum
  of the $\tau$ lepton and light leptons has been applied.
  a) is drawn for the positive $\mu$ scenario and b) for the negative
  case.}
  \label{nt_atlf}
\end{figure}

\begin{figure}
  \includegraphics[width=12cm]{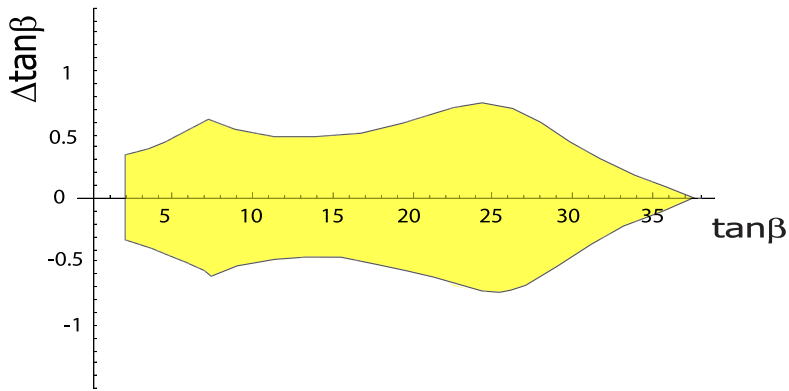}\\
  \caption{The final attainable accuracy on \tanb determination
  for various possible \tanb values. The errors are statistical
  only and evaluated for luminosity of 4 $fb^{-1}$. The plot is done
  for the $\mu>0$ case. The shaded light-blue region represents the
  final result. Note that no ambiguity is left.}
  \label{fin_pos}
\end{figure}

\begin{figure}
  \includegraphics[width=12cm]{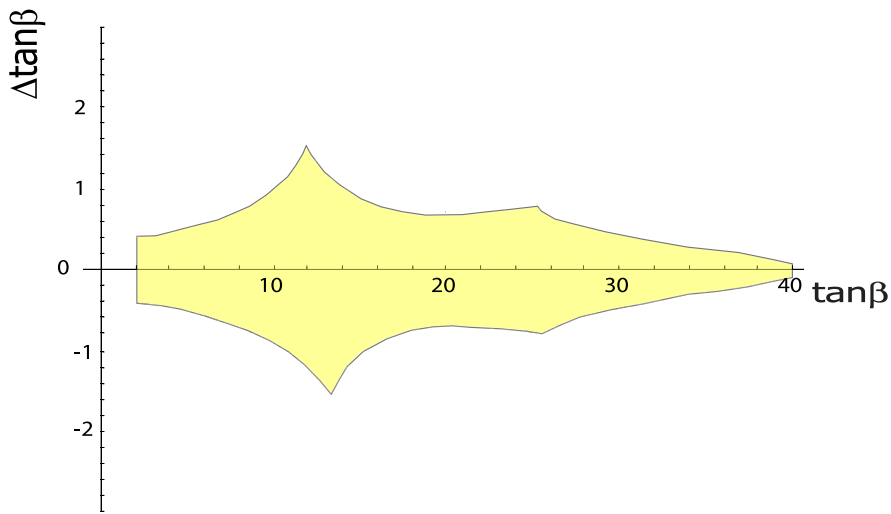}\\
  \caption{The final attainable accuracy on \tanb determination
  for various possible \tanb values. The errors are statistical
  only and evaluated for luminosity of 4 $fb^{-1}$. The plot is done
  for the $\mu<0$ case. Note that no ambiguity is left.}
  \label{fin_neg}
\end{figure}

\begin{figure}
  \includegraphics[width=12cm]{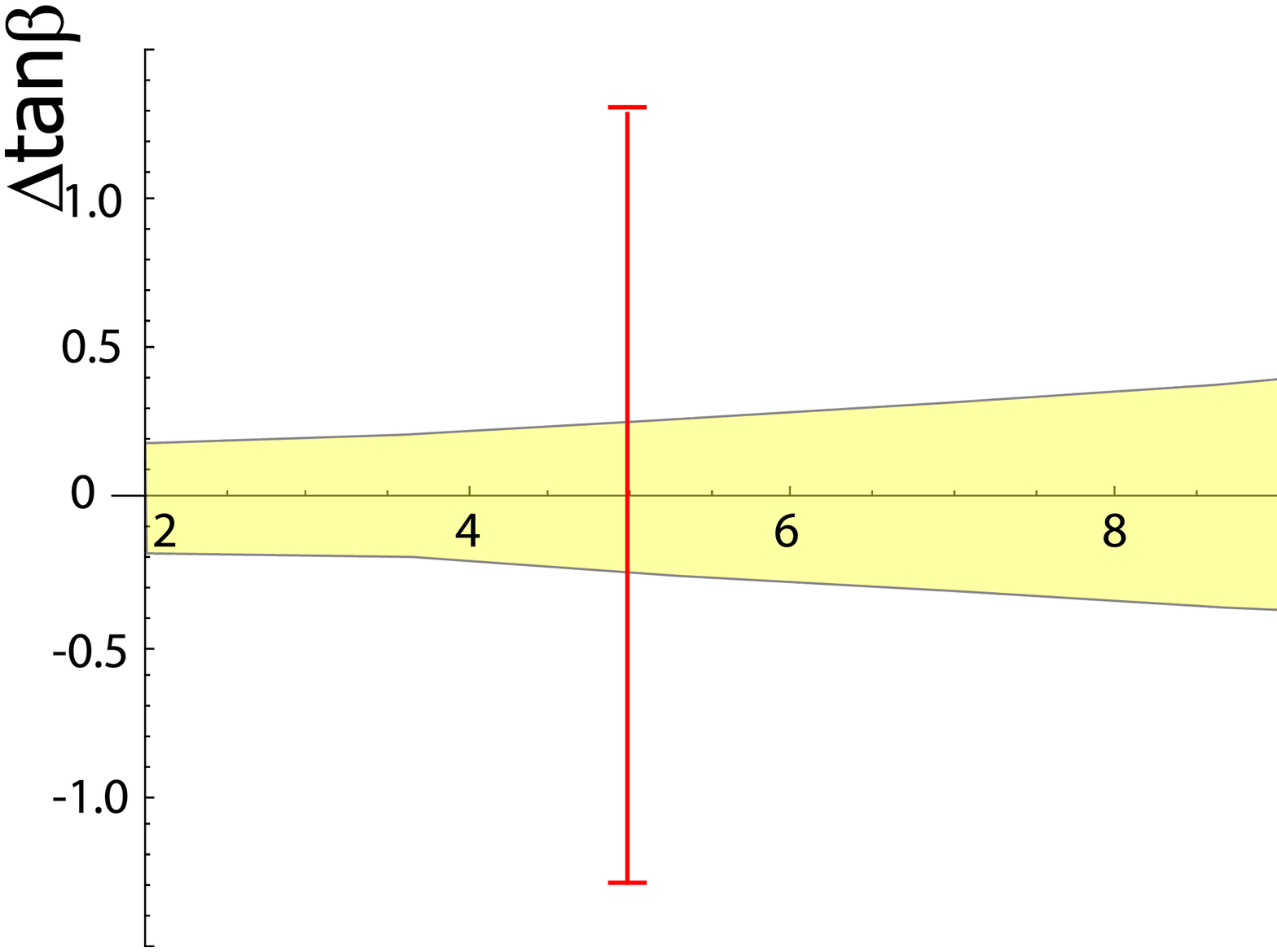}\\
  \caption{A comparison between the present procedure (yellow band)
  and the one based on mass reconstruction \cite{TDR} (red arrow bar
  at \tanbz=5). The present procedure gives a much lower error
  ($\pm 0.25$ compared to $\pm 1.3$) for the same luminosity
  (10 $fb^{-1}$.)}
  \label{comp_2}
\end{figure}

The number of identified $\tau$ leptons and light leptons  is
shown in Figure \ref{nt_atlf} for a luminosity of 4 $fb^{-1}$ for
the case of positive $\mu$ (a) and negative $\mu$ (b). The drop in
the number of leptons at low values of \tanb for the negative
$\mu$ case results from the rise in the $\chi_2^0$ and
$\chi_1^{\pm}$ mass in this region, which is absent in the $\mu>0$
case. By simply counting the number of $\tau$ leptons and light
leptons one can extract the value of \tanb with a small
statistical error as shown in Figures \ref{fin_pos} and
\ref{fin_neg}, in which the error on the reconstructed \tanb is
shown as a function of the input value of \tanb (statistical
errors only). One can also see that the number of light leptons
differs significantly between the positive and negative $\mu$
scenarios as long as $tan\beta<15$. Hence, light lepton counting
will allow a determination of the sign of $\mu$ in this region.
Finally, the error on \tanb was evaluated using the present
procedure under the conditions specified in \cite{TDR}, namely, a
luminosity of 10$fb^{-1}$, M=500 TeV and $\Lambda$= 90 TeV. The
evaluation of the value of \tanb in two independent ways will
provide an important consistency check. The performance of both
methods is compared in Figure \ref{comp_2} with the attainable
performance based on mass reconstruction and the present method is
shown to perform much better.

\section{Systematics}
Simple particle counting is sensitive to uncertainties in the
luminosity measurement as well as  to the uncertainties of the
identification efficiency and purity of the relevant counted
object. The uncertainty in the determination of the luminosity
might be fairly large and might lead to large systematic error in
the determination of \tanb. However, since the proposed procedure
is valid in the framework of a given model which gives rise to a
constant number of photons which does not depend on the value of
\tanb, one would be able to replace the uncertainty on the
luminosity by the statistical error on the number of observed
photons, which is expected to be small due to the abundance of
photons. This 'photon-recalibration' will render the luminosity
error negligible.

\begin{figure}
  \includegraphics[width=10cm]{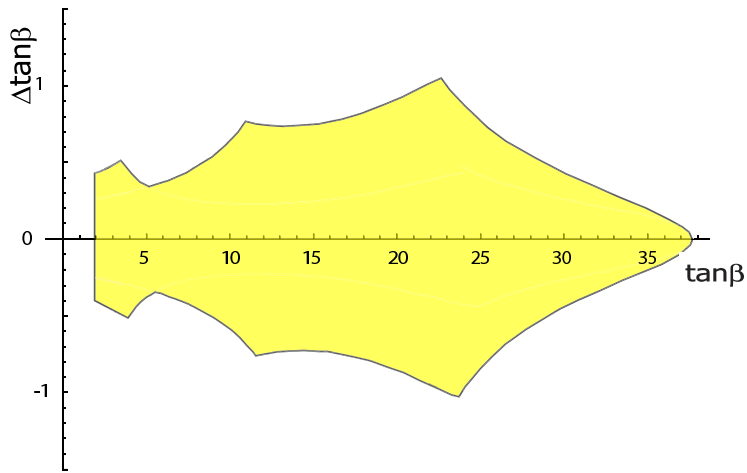}\\
  \caption{Effect of 1\% systematic error on both the detection
  efficiency of light and $\tau$ leptons. Comparing the uncertainty
  in this plot, which is done for positive $\mu$ with Figure
  \ref{fin_pos}  shows the relatively small effect of such systematic
  uncertainty.}
  \label{err_pos}
\end{figure}

The identification efficiency of electrons, muons and $\tau$
leptons and the purity of these samples should be understood, once
$\approx 1fb^{-1}$ has been collected, at a level of ~1\% or
better. This can be done by studying the large number of $Z^0$ and
other SM processes. In order to evaluate the sensitivity of the
present proposed procedure to identification uncertainties a 1\%
shift in both light and $\tau$ lepton identification was
introduced and the value of \tanb was recomputed and is shown in
Figure \ref{err_pos}. As can be seen by comparing this to the
result shown in Figure \ref{fin_pos} in which no systematic error
is introduce, the effect of 1\% systematic error in the
identification efficiency is quite limited.

\section{Conclusion}
A simple electron, muon and $\tau$ leptons as well as photons
counting is shown to provide a fairly accurate determination of
\tanb in N=1 SUSY GMSB case. The proposed technique does not
depend on the measurement of the Higgs boson or $\tilde \tau$ mass
and hence, can be carried out with low luminosity, well before the
Higgs boson mass can be measured. The attainable accuracy is
significantly better than the one obtained by the conventional way
\cite{TDR} of using Higgs boson mass and by explicitly
reconstructing the $\tilde \tau$ lepton.

\end{document}